\documentclass[10pt, conference, compsocconf]{IEEEtran}
\ifCLASSINFOpdf
\else
  \usepackage[dvips]{graphicx}
\fi
\begin{document}
%
\title{Methods of QoS improvement for P2P IPTV based on traffic modelling}


\author{\IEEEauthorblockN{Arkadiusz Biernacki}
\IEEEauthorblockA{Institute of Computer Science\\
Silesian University of Technology\\
Gliwice, Poland\\
Email: arkadiusz.biernacki@gmail.com}
}


%


\maketitle

\begin{abstract}
Over the last years many technological advances were introduced in Internet television to meet user needs and expectations. However due to an overwhelming bandwidth requirements traditional IP-based television service based on simple client-server approach remains restricted to small group of clients. In such situation the use of the peer-to-peer overlay paradigm to deliver live television on the Internet is gaining increasing attention. Unfortunately the current Internet infrastructure provides only best effort services for this kind of applications and do not offer quality of service.

This paper is a research proposition which presents potential solutions for efficient IPTV streaming over P2P networks. We assume that the solutions will not directly modify existing P2P IPTV protocols but rather will be dedicated for a network engineer or an Internet service provider which will be able to introduce and configure the proposed mechanisms in network routers. 

\end{abstract}


%
\IEEEpeerreviewmaketitle

\section{Introduction}
Television is one of the most dominant and pervasive mass media; it is watched across all age groups and by almost all countries in the world. Over the last years, many technological advances were produced by trying to meet user needs and expectations in such a widespread media. Traditional Internet TV (IPTV) service based on simple client-server approach restricted small group of clients, the overwhelming bandwidth requirement makes it impossible when the number of user grows to thousands or millions because servers have limited available resources (CPU, bandwidth) that will decrease proportionally with the number of users. By multiplying the servers and creating content distribution network (CDN), the solution will scale only to a larger audience with regards to the number of deployed servers which may be limited by the infrastructure costs. Finally, the lack of deployment of IP-Multicast limits the availability and scope of this approach for a TV service on the Internet scale. Therefore the use of the peer-to-peer overlay paradigm (P2P) to deliver live television on the Internet (P2P IPTV) is gaining increasing attention, and has become a promising alternative \cite{liu_opportunities_2008}⁠.

\begin{figure}[!t]
\centering
\includegraphics[width=3.1in]{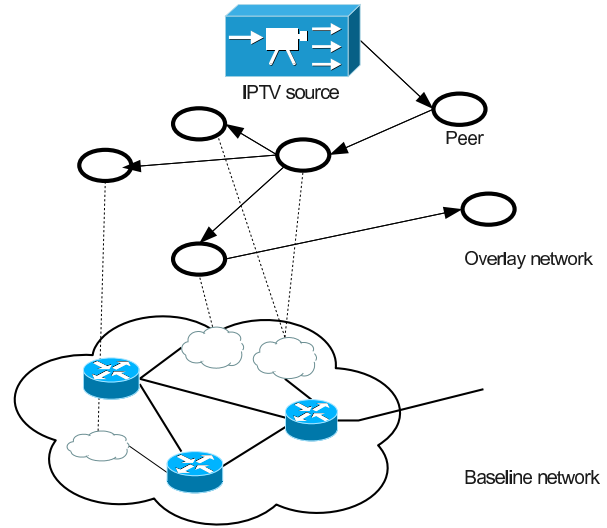}
\caption{P2P IPTV system}
\label{fig:camera}
\end{figure}

From the technical point of view, in P2P IPTV system, a local user (or peer) act both as receiver and supplier of the IPTV data. Connected to “upstream” pears, the user receives IPTV data chunks and forwards them to other “downstream” receivers, watching the program which, in turn, can forward it further down in a hierarchy of peers -- fig. \ref{fig:camera}. Consequently, such an approach has the potential to scale with group size, as greater demand also generates more resources. Moreover, by using the existing Internet infrastructure as a medium and by exploiting user participation for the creation of the content distribution network, P2P IPTV technologies have innovative potential by making any TV channel from any country globally available and allowing Internet users to broadcasting their own TV with low costs.

\section{P2P research challenges}
P2P applications are posing serious challenges to Internet infrastructures and there is continuous battle between service providers and P2P applications for traffic management. One of the major problems is QoS provisioning. Unlike file sharing, the live media need to be delivered almost synchronously to large number of users, with minimum delay in playback compared to the playback at the source. Due to the large volume of data in the media stream, it is of paramount interest to avoid redundant transmission of the stream. Constructing efficient paths for streaming is especially hard because the nodes participating in the overlay have very minimal information regarding the topology of the baseline network.

Real time services such as IPTV are inelastic, as the transmission bandwidth, transmission time and QoS requirements need to be kept within strict limits and hence are not flexible. Current Internet infrastructure provides best effort services that do not offer quality of service. These issues result into lower throughput (bandwidth management), packet losses, high transfer delay, delay variation (jitter) and out-of-order delivery. These parameters are unpredictable and never acceptable for real-time applications. Thus, we need to design solutions for efficient video streaming over P2P networks that can address the above-mentioned issues. We assume that the solutions will not directly modify existing P2P IPTV protocols but rather will be dedicated for network engineer or Internet service provider (ISP) which will be able to introduce and configure the proposed mechanisms in network routers.
 
In the prosed research we will try to answer the following questions: in which case proper ISP assistance is needed to help P2P IPTV system and what are the most proper services ISP should provide to support P2P IPTV. Taking into account the above assumptions we will focus on QoS provisioning for P2P IPTV services by improving the overall received video throughput with low packet drop ratio, transmission delay and jitter -- treating them simultaneously as our QoS metrics. We concentrate on provisioning to P2P IPTV some of the methods influencing QoS based on differentiated services and including traffic control which regulates data flows by classifying, scheduling, shaping traffic and admission control determining which applications and users are entitled to network resources. 

\begin{figure}[!t]
\centering
\includegraphics[width=3.1in]{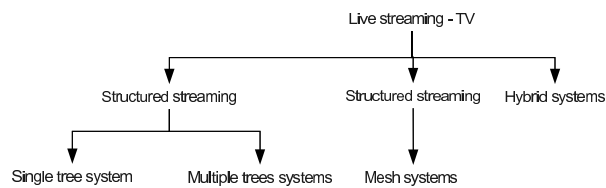}
\caption{Classification of {P2P} streaming systems}
\label{fig:classyfication1}
\end{figure}

There are also other mechanisms which lead to improvement of QoS. One of them is redesign overlay network topology which should results in improving proposed QoS metrics. Early P2P streaming systems were designed as alternatives to IP multicast. Therefore they all attempt to build application level multicast trees. However, a tree structure is unsuited in a P2P environment. A tree is vulnerable to node failures and streaming rate of a peer cannot exceed that of its parent. To address the problem of tree based architectures, current systems adopts rather a multi-parent mesh approach for P2P streaming -- fig. \ref{fig:classyfication1}. The idea is to allow each peer to stream media data from multiple neighbour peers. To coordinate the streaming from multiple sources in P2P IPTV system, usually a pull-based approach is used where peer collects data availability from its neighbours, and request different data blocks from different neighbours. While the multi-parent, receiver-driven approach offers great flexibility in dealing with peer and network dynamics, a remaining question is how should peers select their neighbours and what does the resulting overlay look like? P2P applications usually do not implement any algorithms to select peers with the best connectivity (e.g. considering delays, throughput, etc.). There exists solutions improving the connection choice based on hybrids mixing a tree with mesh structure \cite{zhang_peer-to-peer_2005}⁠ or peer ranking \cite{xie_p4p:_2008}⁠\cite{aggarwal_isp-aided_2008} but in most of the P2P streaming platforms the overlay neighbours are chosen randomly \cite{zhang_coolstreaming/donet:data-driven_2005} \cite{wu_magellan:_2007}⁠.

\section{QoS improvement propositions}
Taking into account the goals stated in previous section we proposed a few solutions which should improve the overlay topology between peers and should help deliver video content efficiently. Proposing the solution we assume that ISPs and P2P systems collaborate so that both benefit: 1) the ISP keeps a significant portion of their network traffic localised within their internal network, and hence gain cost advantages by reducing costs for traffic that leaves their network boundary 2) better management of traffic flow may provide better service to customers and ensure fairness for other applications like VoIP or web traffic. 

\subsection{Caching}
\begin{figure}[!t]
\centering
\includegraphics[width=3.1in]{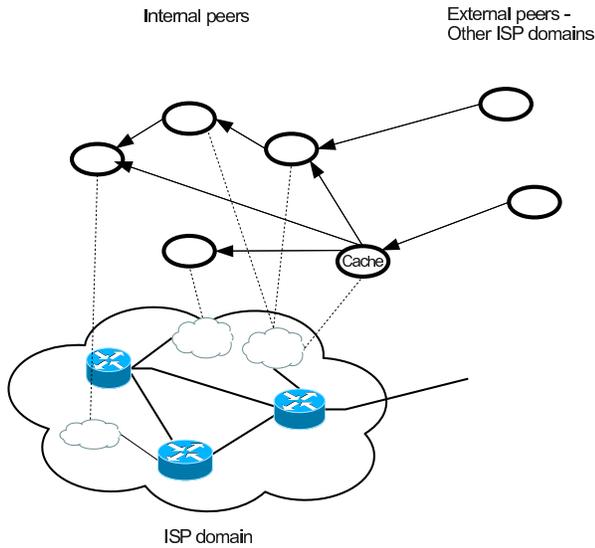}
\caption{Cache server placement problem}
\label{fig:p2pCache1}
\end{figure}

P2P caching is similar in principle to the content caching long used by ISPs to accelerate Web (HTTP) content. P2P caching temporarily stores popular content that is flowing into an ISP’s network. If the content requested by a subscriber is available from a cache, the cache satisfies the request from its temporary storage, eliminating data transfer through expensive transit links and reducing network congestion. However we must note that in case of IPTV the expiration time of the cache content will be quite short in contrary to other P2P services like file sharing. Content placement deals with how many replicas of each object has and where in the network to place them. Intuitively, cache servers should be placed in such a manner that they are closer to the clients, thereby reducing latency and bandwidth consumption -- fig. \ref{fig:p2pCache1}. Also, content replicas should be placed to even the load of the replica servers in P2P network, that is, trying to balance the load among cache servers. In a our case, TV content from source server is delivered to P2P network which is responsible for streaming the content between its peers. So one specific question is: how to select from the available peers a streaming server within the given ISP domain? Another question is: how to place cache servers among all possible locations to cut down cost and improve performance of the whole ISP network? After selecting the location of servers, we need to decide what amount of media content the server hold. The problem of server placement may be modelled as center placement problem: for the placement of a given number of centers, minimise the maximum distance between a node and the nearest center. There are several class of algorithms which are used for solving this problem: greedy, hot-spot and tree-based \cite{weigmann_center_1997}⁠. In the particular case a cache server be may installed at the border between the local user base of the ISP and the Internet cloud. Based on destination port number for each TCP connection all P2P IPTV traffic can be redirected to this server. Thus, the server is able to intercept all downloads performed by local users \cite{wierzbicki_cache_2004}.
The aforementioned idea is presented in literature in context of CDN and P2P networks. In \cite{shen_hptp:_2007}⁠ authors proposed to use caching to relieve the tension between ISPs and P2P systems. A network layer packet-level caching for reducing the volume of emerging P2P traffic is proposed in \cite{nakao_remedy_2008}⁠. In \cite{chen_server_2008} authors formulate server placement problem and in P2P streaming system and propose solution schemes for the sub-problem of server selection and rate assignments.

\subsection{Prioritisation and smoothing}
One method to reduce information loss (due to buffer overflow or end-to-end delay) experienced by multimedia traffic is through the use of multiple priority traffic classes within the network. Assigning higher priority to multimedia traffic throughout the network will prevent latency tolerant data traffic from delaying time critical multimedia streams. Traffic priority classes will also improve performance in systems whose peak data rate exceeds network capacity. 

\begin{figure}[!t]
\centering
\includegraphics[width=3.1in]{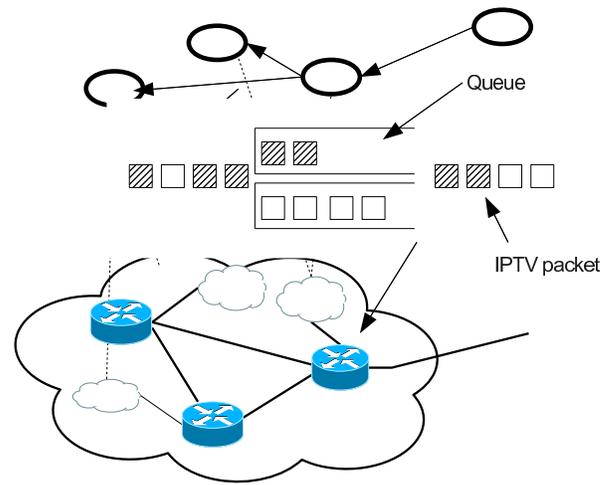}
\caption{P2P IPTV packets scheduling}
\label{fig:scheduling}
\end{figure}

When there are large bursts of data traffic, there will be severe congestion on the network. The end-to-end delay experienced by the multimedia streams will increase, resulting in poor performance. Traffic flows are aggregated in the network, so that core routers only need to distinguish a comparably small number of aggregated flows, even if those flows contain thousands or millions of individual flows. In order to support real time traffic, we need a mechanism to prioritise data. This is done by classifying traffic into service classes based on expected traffic patterns. Each service class has a data priority level and associated guarantees.  Scheduling algorithms determine which packet to send next and are used primarily to manage the allocation of bandwidth between flows. We propose this approach for P2P IPTV traffic which needs real time guarantees -- fig. \ref{fig:scheduling}. Applying different scheduling algorithms we intend to observe the dynamic and interaction of overlay with baseline network. For example the simplest scheduling is always transmitting high priority traffic first. If the multimedia traffic is given priority, it will not be affected by these large bursts of data.  This guarantees the IPTV traffic will always have good performance. However, it is possible that lower priority traffic will never be serviced. This will be true as long as the average bandwidth requirements do not exceed the capacity of the network. Once the network capacity becomes insufficient to handle the average requirements, the only solutions are to increase capacity, change scheduling algorithm or impose restrictions on the transmitting stations. Taking into account the last case we will attempt to use queue management algorithm to prevent network overloading.

\begin{figure}[!t]
\centering
\includegraphics[width=3.1in]{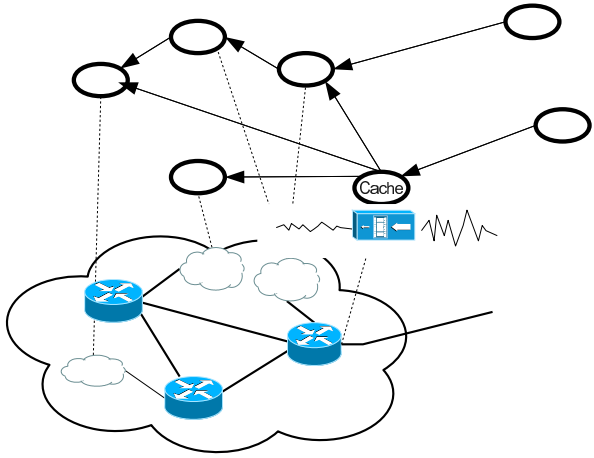}
\caption{P2P IPTV traffic smoothing}
\label{fig:smoothing}
\end{figure}

Compressed digital video is inherently variable-rate since its complexity and motion content affect the encoding bit rate required to maintain picture quality which leads to uncontrolled burstiness. Traffic burstiness is one of the reasons of inefficient use of network resources by occasionally requiring excessively high processing, storage (buffering), and transmission capacity from the network. By transmitting smoother traffic, the network can improve its utilisation. However the price for smoothing is either reduction of bandwidth the streams consumes or packet drops or delays in stream transmission. We propose delaying the playback of the P2P IPTV traffic to permit the source a transmission over a larger interval (window) of frames, based on the buffer space available at the client site. However even if we delay the stream transmission, if such smoother stream traverses multiple hops its queuing delay is reduced on each hop resulting in considerable decrease in whole end-to-end delay bound which overweights the initial delay \cite{paola_improving_1996}⁠. We propose to implement traffic shapers in cache servers mentioned in previous proposition or in border router. The traffic shapers may be switched on and off depending on the hop distance between the cache server and a receiver -- fig. \ref{fig:smoothing}.

\subsection{Congestion avoidance}
\begin{figure}[!t]
\centering
\includegraphics[width=3.1in]{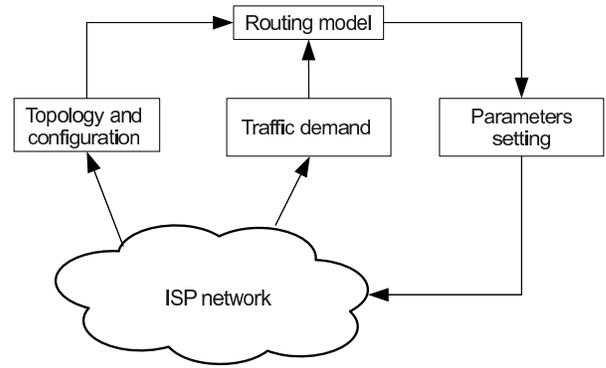}
\caption{ISP routing improvement in presence of P2P IPTV traffic}
\label{fig:routing}
\end{figure}

As it was stated in previous section in most P2P streaming platforms, the overlay neighbours are chosen randomly which may lead to the situation when one link may be heavily congested while other links in the network remain lightly loaded or the P2P IPTV packets may traverse a long path with high propagation delay when a low-latency path is available. The responsibility of selecting the path, which a packet follows through the baseline network, falls to the routing protocols implemented by the individual routers in the network. Rather than using hard-wired tables to forward the packets, the routers exchange control messages with each other to compute the paths through the network in a distributed fashion. The distributed approach allows a collection of routers to adapt automatically to changes in the network topology. This makes IP networks robust in the presence of link and router failures, and easily accommodates the deployment of new equipment as the network grows. However, the routing protocols deployed in most IP networks do not incorporate information about network load and performance into the selection of the paths. Left to their own devices, the routers continue to forward packets over heavily loaded link. A near-optimal intra-domain load balancing can be achieved by altering the link weights of the network’s routing protocol (OSPF or IS-IS) \cite{rexford_route_2005}. The selection of weights depends on having an estimate of the offered load on the network, in terms of the volume of traffic between each pair of routers or each pair of edge links. This kind of information is necessary for designing the network and planning the outlay of new capacity. In some cases, the operator may have an estimate of the traffic demand based on past experience or customer subscription information. In other cases, the traffic demands can be obtained by measuring the traffic in the operational network. Computing estimates of the offered load requires combining measurement data from multiple locations in the network to compute the traffic demands. This allows the use of optimisation techniques for identifying parameter settings that satisfy the network's performance goals -- fig. \ref{fig:routing}.

\section{Implementation draft}
Proposed traffic engineering tasks including caching, scheduling, smoothing and load balancing require an effective way to predict the flow of traffic through the network. In such circumstances traffic matrices (TM) are helpful which reflect the volume of traffic that flows between all possible pairs of sources and destinations in a network. Constructing a network-wide view of the traffic demands requires relatively sophisticated techniques for the collection and analysis of measurement data. Traffic statistics may be available directly from SNMP, can be computed by combining packet-level or flow-level measurements at the network edge with the information available in routing tables, may be inferred based on observations or sampling of the aggregate load on links inside the network in conjunction with routing data \cite{fortz_traffic_2002}⁠. 

\begin{figure}[!t]
\centering
\includegraphics[width=3.1in]{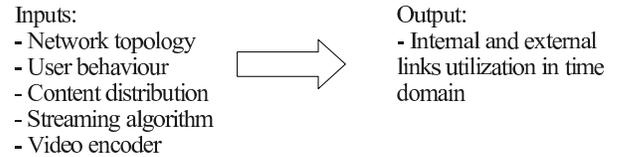}
\caption{Inputs and output for the proposed traffic model}
\label{fig:trafficModel}
\end{figure}

The above methods of TM estimation would require access to ISP network equipment. Additionally the measurement would present a snapshot of ISP traffic in certain period of time with no access to parametrisation like 1) dynamics of the P2P system – nodes availability; 2) nodes links capacity; 3) TV channel popularity; 4) overlay and baseline networks topology; 5) exchange protocol – e.g. peer selection, chunks scheduling; 6) video encoder used. Instead we propose to use physical modelling for P2P network traffic which tries to explicate the physical causes of certain traffic behaviour in the system based on network mechanisms and empirical established properties of the system. In order to build a comprehensive P2P network traffic model we must take into account behaviour  of P2P system in three main categories: individual peer behaviour, shared contents (TV) and multiple peer characteristics. As a result of the model, we obtain internal and external ISP link utilisation in time domain -- fig. \ref{fig:trafficModel}.

Model parameters will be obtained from analysis and measurement of real open-source P2P IPTV application GoalBit \cite{bertinat_goalbit:first_2009}⁠. GoalBit is capable of distributing high-bandwidth live-content using a Bittorrent-like approach where the stream is decomposed into several flows sent by different peers to each client. The system has also built-in mechanism of perceived quality measurement.
P2P traffic can be broadly classified into two categories: signalling and data transfer. The signalling traffic includes TCP connection set-up, search queries and query replies. Its volume heavily depends on the type of P2P network (structured or unstructured) and protocol used. The leading content shared in the P2P IPTV systems tend to be larger in size compared to the signalling traffic although the signalling traffic packets are sent more frequently \cite{silverston_traffic_2009}⁠. To obtain comprehensive view of network behaviour we will take into account both types of traffic. 

The traffic model will be important element of our project and will be applied in all proposed solutions: 1) traffic caching -- due to the spatio-temporal traffic model we may identify the optimal placement of the cache server; 2) traffic scheduling -- the temporal traffic model will be used for selection of optimal parameters for scheduling algorithm \cite{lazar_modeling_1994}⁠; 3) traffic smoothing -- temporal traffic model will allow to choose optimal parameters for the smoothing mechanisms; 4) weights manipulation -- the spatial traffic model may be used instead of the costly and time intensive measurements of traffic intensity in ISP network.

The next issue is the implementation and comparison of the traffic models and proposed solutions. We take into account analytical models, simulations and experiments on the real systems. Large-scale distributed systems are complex and accurately modelling them analytically is not an easy task. In many cases the first iteration of an analytical model is not tractable and the model needs to be successively simplified to produce useful insights. This leads to models that describe the system at a very coarse-grained level and with many uniformity assumptions. Despite this, we plan to use analytical models for early feasibility assessment of a P2P solutions. However, for a more complete and accurate evaluation under a wider range of conditions we incline towards a simulation and system evaluation on the real networks. 

Usually the simulators are purpose-built for specific applications or classes of applications. Few simulators are designed as more general tools for system building and evaluation. What is more, even though simulation is a widely used evaluation technique there is almost no simulator code sharing among the researchers and little standardisation of the common practises \cite{naicken_survey_2006-1}. There are a number of dedicated P2P simulators although no all of them have the functionality that we would expect. Given these issues with current P2P simulators and the importance of reproducing the results, in this project we plan use open-source OMNeT++. In order to obtain spatio-temporal traffic model the simulation of overlay network will be based on underlying network including most important physical network elements like workstations, routers, links etc. 

Validation of a model of complex P2P system based on performing experiments with the actual system requires significant resources which could be very costly in hardware and administration, and is vulnerable to node failures. There may also be factors external to the experiment that can not be controlled, yet influence experiment results, such as cross-traffic and changes in the properties of the baseline network. Thus we plan to use where possible a virtual test-bed -- Planetlab \cite{chun_planetlab:overlay_2003}⁠ or P2P-Next \cite{jimenez_p2p-next:_2009}⁠ which are useful tool for performing large scale experiments on overlay protocols and for validating some results obtained from simulation results. For creation our P2P IPTV application we consider to use a peer-to-peer systems prototyping toolkit ProtoPeer that allows for switching between the event-driven simulation and live network deployment without changing any of the application code ⁠\cite{galuba_protopeer:simulation_2008}.

For the most of the proposed solutions the P2P IPTV traffic flows must first be identified. We assume that popular traffic identification techniques are available based for example on ports numbers or signature matching. We also consider applying more advanced method of statistical traffic identification based on the developed traffic model and guidance presented in \cite{kim_internet_2008}⁠.

\section{Conclusions}
In the paper we proposed a few solutions which should improve QoS in P2P IPTV network. We assume that the solutions will not directly modify existing P2P IPTV protocols but rather will be dedicated for network engineer or Internet service provider which will be able to introduce and configure the proposed mechanisms in network routers. The proposed methods require an effective way to predict the flow of traffic through the network. We are going to use physical modelling for P2P IPTV network traffic which tries to explicate the physical causes of certain traffic behaviour in the system based on network mechanisms and empirical established properties of the system. The challenge lies in combining relevant feature from workload modelling, network architecture (topology, protocols), users behaviour, and analytical modelling into consistent description of P2P system. Because we are to model a complex distributed system we incline towards a simulation and system evaluation on the real networks. We expect to find answer for the question: which of the proposed solutions is most suitable for QoS improvements in P2P IPTV networks taking into account its efficiency, easiness of implementation and potential side effect on other kinds of network traffic.

\bibliographystyle{IEEEtran}
\bibliography{zotero}
\end{document}